\documentclass[twocolumn,aps,prl,groupedaddress]{revtex4}
\usepackage{graphicx}
\usepackage{color}
\usepackage{amsmath}
\usepackage{enumitem}

\newcommand{\be}{\begin{equation}}
\newcommand{\ee}{\end{equation}}

\newcommand{\bea}{\begin{eqnarray}}
\newcommand{\eea}{\end{eqnarray}}

\newcommand{\p}{\partial}

\newcommand{\la}{\left\langle}
\newcommand{\ra}{\right\rangle}
\newcommand{\lb}{\left[}
\newcommand{\rb}{\right]}
\newcommand{\lp}{\left(}
\newcommand{\rp}{\right)}

\renewcommand{\Im}{{\rm \, Im\,}}
\renewcommand{\vec}[1]{{\bf #1}}
\renewcommand{\phi}{\varphi}
\renewcommand{\epsilon}{\varepsilon}

\newcommand{\addLL}[1]{\textcolor{blue}{#1}}

\begin{document}
\title{Tomographic Dynamics and 
Scale-Dependent Viscosity in 
2D Electron Systems} 

\author{Patrick Ledwith${}^1$, Haoyu Guo${}^1$, 
Andrey Shytov${}^2$, Leonid Levitov${}^1$} 
\affiliation{$^1$Massachusetts Institute of Technology, Cambridge, Massachusetts 02139, USA \\ $^2$ School of Physics, University of Exeter, Stocker Road, Exeter EX4 4QL, United Kingdom}



\begin{abstract} 
Fermi gases in two dimensions  display a surprising collective behavior originating from the head-on carrier collisions. The head-on processes dominate 
angular relaxation at not-too-high temperatures $T\ll T_F$ owing to the interplay of Pauli blocking and momentum conservation. As a result, a large family of excitations emerges, associated with the odd-parity harmonics of momentum distribution and having exceptionally long lifetimes. This leads to ``tomographic'' dynamics: fast 1D spatial diffusion along the unchanging  velocity direction accompanied by a slow angular dynamics that gradually randomizes velocity orientation. The tomographic regime features an unusual hierarchy of time scales and scale-dependent transport coefficients with nontrivial fractional scaling dimensions, leading to fractional-power current flow profiles and unusual conductance scaling vs. sample width.
\end{abstract}

\maketitle
Electron transport 
in many systems of current interest is 
governed by the processes of rapid momentum exchange in carrier 
collisions\cite{dejong_molenkamp,bandurin2015,crossno2016,moll2016}. 
Disorder-free electron systems, in which 
the  
electron-electron (ee) collisions are predominantly momentum-conserving, 
can exhibit a hydrodynamic behavior reminiscent of that in viscous fluids 
\cite{gurzhi63,LifshitzPitaevsky_Kinetics,jaggi91,damle97}.
Electron hydrodynamics,
a theoretical concept describing this regime in terms of quasiparticle scattering near the Fermi surface, has been steadily gaining 
 support in recent years 
 \cite{andreev2011,sheehy2007,fritz2008,muller2009, 
 forcella2014,tomadin2014,narozhny2015,principi2015}. 

\begin{figure}
\includegraphics[scale=0.4]{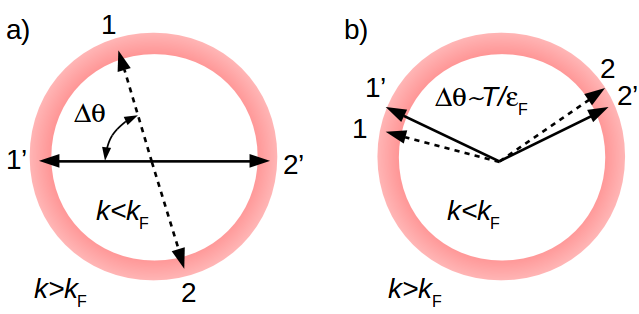}
\caption{ 
Types of two-body collisions $1,2\to 1',2'$ 
at a thermally broadened 2D Fermi surface (red rings), which are allowed by momentum and energy conservation and not inhibited by fermion exclusion.  
Head-on collisions (a) 
occur at a rate $\gamma\sim T^2/T_{\rm F}$, 
with typical recoil $\Delta\theta\sim 1$. 
Such processes, however, affect only the even-parity part of momentum 
distribution. 
The odd-parity part, in contrast, relaxes solely due to small-angle scattering (b). 
Angular diffusion with a step $\Delta\theta\sim T/T_{\rm F}\ll1$ 
slows down the odd-parity relaxation, 
reducing the relaxation rate  down to 
$\gamma'\sim T^4/T_{\rm F}^3\ll\gamma$.
}
\label{fig1}
\vspace{-5mm}
\end{figure}

It is usually taken for granted that hydrodynamics sets in at the lengthscales 
$r>l_{\rm ee}=v/\gamma$ 
where  $\gamma\sim T^2/T_{\rm F}$ is the ee collision  rate
and $v$ is Fermi velocity.
Here we argue that in 2D systems---the focus of current experimental efforts\cite{dejong_molenkamp,bandurin2015,crossno2016,moll2016}---our understanding 
of electron hydrodynamics requires a substantial revision. 
Indeed, generic large-angle quasiparticle scattering at a thermally broadened 2D Fermi surface is inhibited by fermion exclusion, except for the head-on scattering, which dominates angular relaxation (see Fig.\ref{fig1}) \cite{laikhtman_headon,gurzhi_headon,molenkamp_headon}. 
The head-on collisions do lead to rapid momentum exchange between particles,  however with one caveat. 
Such collisions change particle 
distribution 
in an identical way
at momenta $\vec p$ and $-\vec p$,
providing relaxation pathway only for the part of momentum 
distribution which is {\it even} under  Fermi surface inversion, $\delta f_{-\vec p}=\delta f_{\vec p}$. The odd-parity part $\delta f_{-\vec p}=-\delta f_{\vec p}$ does not relax due to such processes, giving rise to a large number of soft modes \cite{SOM}. This peculiar behavior is generic in 2D at $T\ll T_F$, so long as the ee collisions are momentum-conserving. 

The new 
regime, dominated by the head-on collisions and odd-parity harmonics, 
occurs at the lengthscales  (and frequencies) in
between the conventional ballistic and hydrodynamic regimes, 
\be\label{eq:lengthscales}
l_{\rm ee}<r<\xi=\frac{v}{\sqrt{\gamma'\gamma}}
,
\ee
where 
$\xi\gg l_{\rm ee}$ is a new lengthscale originating from slowly relaxing odd-parity modes. 
Here the rate  $\gamma\sim T^2/T_{\rm F}$ describes head-on 
processes and even-parity modes, 
the rate $\gamma'\ll\gamma$ describes 
slow odd-parity modes. 
The intrinsic $\gamma'$ values due to small-angle ee scattering 
are estimated to be as low as \cite{SOM}
\be\label{eq:gamma'}
\gamma'\sim (T/T_{\rm F})^2\gamma\ll \gamma
.
\ee
Since the rate $\gamma'$ is small, in real systems it may be overwhelmed by extrinsic effects, 
such as phonons or disorder.

The conventional ballistic and hydrodynamic regimes occur at
$r<l_{\rm ee}$ and $r>\xi\gg l_{\rm ee}$, respectively. 
In the ballistic regime 
the system features a standard free-particle behavior. 
Likewise, in the hydrodynamic regime 
transport coefficients assume their conventional values, e.g. the standard result $\nu=v^2/4\gamma$ for kinematic viscosity. However, at the intermediate scales \eqref{eq:lengthscales} transport coefficients acquire a 
dependence on the wavenumber, becoming scale-dependent with nontrivial 
scaling dimensions. 

At this point one may ask why the quasiparticle lifetimes, evaluated from the many-body Green's functions as $\frac1{2\tau}=\Im\Sigma(\epsilon,p)$,  behave as $T^{-2}/\ln\frac{T_{\rm F}}{T}$ (or, $\epsilon^{-2}/\ln\frac{\epsilon_{\rm F}}{\epsilon}$ at zero temperature) without showing any indication of the slow modes\cite{chaplik71,hodges1971,bloom1975,giuliani82,chubukov2003}. This is so because the lifetimes evaluated by the selfenergy method are dominated by the fastest decay pathway, with the slow pathways due to long-lived modes providing a subleading contribution to the decay rates. A different scheme is therefore required for treating the slow and fast modes on equal footing. 

Here we consider a simple model 
in which 
different harmonics of particle momentum distribution $\delta f_{\vec p}=\sum_m \delta f_m e^{im\theta}$, with the angle $\theta$ 
parameterizing the Fermi surface,  
relax at different rates. We will assume that the even-$m$ harmonics relax at a 
constant rate $\gamma\sim T^2/T_{\rm F}$, whereas the odd-$m$  
rates behave as $\gamma'm^p$ with $\gamma'\ll\gamma$:
\be\label{eq:I_m}
\gamma_{m\, {\rm even}}=\gamma(1-\delta_{m,0})
,\quad \gamma_{m\,{\rm odd}}=\gamma' m^p (1-\delta_{m,\pm1})
.
\ee
Zero values for $\gamma_{m=0,\pm1}$ reflect particle number and momentum conservation. 

Below we consider several different $p$ values which describe different regimes of interest. 
The intrinsic relaxation mechanism due to ee collisions predicts the odd-parity relaxation with $p=4$ \cite{SOM}. In addition, we consider the cases $p=2$ and $p=0$. This is done for illustration as well as having in mind that,  in real systems, the very long lifetimes due to intrinsic effects can be overwhelmed by extrinsic effects. Relaxation due to residual disorder, phonons or scattering at boundaries 
is described by $p=0$, whereas small-angle scattering due to smooth disorder potential, leading to conventional angular diffusion, 
is described by $p=2$. In these cases, $\gamma'$ is governed by other effects than the ee interactions. 
The intrinsic $m^4$ scaling of the odd-$m$ rates corresponds to angular superdiffusion, with $\gamma'$, given by Eq.\eqref{eq:gamma'}, taking on a role of the angular diffusion coefficient (see Eqs.\eqref{eq: angle_diffusion},\eqref{eq:master_eqn} below). 

It might seem surprising that the modes with high $m$ values could 
impact transport properties, since particle density and current---the two quantities usually probed in experiments---are described by $m=0,\pm1$ harmonics.  
Qualitatively, the significance of the high-$m$ modes 
can be understood on very general grounds in terms of the Fluctuation-Dissipation Theorem which 
mandates strong fluctuations for slowly-relaxing degrees of freedom. 
Strong fluctuations, in turn, translate into enhanced scattering  
for other degrees of freedom, provided those are coupled to the slow degrees of freedom. 

To understand 
how different slow modes are coupled, we consider transport equation, linearized near the $\vec p$-isotropic equilibrium state:
\be\label{eq:Boltzmann_linearized}
(\p_t+\vec v\nabla-I_{\rm ee})\delta f_{\vec p}(t,\vec x)=0
.
\ee 
Couplings between different angular harmonics arise from the $\vec v\nabla$ term. To elucidate these couplings, we transform Eq.\eqref{eq:Boltzmann_linearized} 
to the $\delta f_m$ basis, 
$\delta f_{\vec p}=\sum_m\delta f_m e^{im\theta}$. 
For plane-wave modes $\delta f_{\vec p}(t,\vec x)\sim e^{i\vec k\vec x-i\omega t}$, in the $\delta f_m$ basis Eq.\eqref{eq:Boltzmann_linearized} takes the form of a 
1D tight-binding model in which the eigenvalues of $I_{\rm ee}$ and $\frac{ikv}2$ 
represent the on-site potential and nearest-neighbor hopping amplitudes:
\be\label{eq:tight_binding}
(\gamma_m-i\omega)\delta f_m=\frac{ikv}2 \delta f_{m+1}+\frac{ikv}2 \delta f_{m-1}
\ee
 (without loss of generality we choose $\vec k\parallel x$).
The hopping terms in Eq.\eqref{eq:tight_binding} arise since $\cos\theta f(\theta)$ Fourier-transforms to $\frac12f_{m+1}+\frac12f_{m-1}$. 
For $\gamma_m$ values vanishing on every other site, as in Eq.\eqref{eq:I_m} in the limit $\gamma'/\gamma\to 0$, one can construct a 
non-decaying  ($\omega=0$) Bloch eigenstate described by $\delta f_m$ vanishing on all the decaying sites with $\gamma_m\ne0$ but nonzero and alternating in sign on the non-decaying sites where $\gamma_m=0$, namely
\be\label{eq:dark_state}
\delta f_{m=2s+1}=(-1)^s,\quad \delta f_{m=2s}=0
. 
\ee
Eq.\eqref{eq:dark_state} represents {\it a dark eigenstate} which is infinitely long-lived. 
Furthermore, 
the system hosts an entire 
band of long-lived near-dark states, 
with the lifetimes diverging in proximity of the dark state. 
Since these states have nonzero overlaps with the $m=\pm 1$ harmonics that govern electric current,  slow decay translates---by the fluctuation-dissipation theorem---into an enhancement of current fluctuations and higher conductivity.
The latter, in turn, means reduced dissipation and lower viscosity. 

The essential physics here resembles the slow-mode relaxation mechanism by Mandelshtam and Leontovich, and Debye, with the $m>2$ harmonics playing the role of bath variables (see, e.g., \cite{Levanyuk} and references therein).  Since mode coupling in Eq.\eqref{eq:tight_binding} is proportional to $kv$, the impact of soft modes with high $m$ is stronger at larger $k$. This can be seen as an underlying reason for transport coefficients such as conductivity and viscosity becoming scale-dependent. 

Turning to evaluating transport coefficients, we consider flows induced by an 
in-plane electric field varying as $\vec E(\vec x)=\vec E_{\vec k}\cos\vec k\vec x$. 
Small deviations from equilibrium 
are described by a linearized kinetic equation
\be\label{eq:Boltzmann_linearized_Efield}
(\p_t+\vec v\nabla_{\vec x}-I_{\rm ee})\delta f_{\vec p}(t,\vec x)=-e\vec E(\vec x)\nabla_{\vec p}f_{\vec p}^{(0)}
,
\ee 
where $f_{\vec p}^{(0)}$ is the equilibrium distribution. The perturbed distribution $\delta f_{\vec p}$ is nonzero near the Fermi surface. Below we will focus on the shear flows, described by $\vec E_{\vec k}\perp\vec k$.

Since the even and odd parts of the distribution $\delta f_{\vec p}(t,\vec x)$ relax at very different rates, we employ an adiabatic approximation 
in order to ``integrate out'' the even-parity part and derive a closed-form equation for the odd-parity part. 
We first note that the only term in Eq.\eqref{eq:Boltzmann_linearized_Efield} that alters parity, $\vec v\nabla_{\vec x}$, transforms functions of odd parity to those of even parity, and vice versa. We can therefore decompose the distribution into a sum of an odd and an even contribution,  $\delta f_{\vec p}=
\delta f_{\vec p}^{+}+\delta f_{\vec p}^{-}$, and write a system of coupled equations for these quantities: 
\bea
&&(\p_t-I_+)\delta f_{\vec p}^{+}(t,\vec x)
+\vec v\nabla_{\vec x}\delta f_{\vec p}^{-}(t,\vec x)=0
,\quad
\\
&& \nonumber
(\p_t-I_-)\delta f_{\vec p}^{-}(t,\vec x)
+\vec v\nabla_{\vec x}\delta f_{\vec p}^{+}(t,\vec x)
=-e\vec E(\vec x)\nabla_{\vec p}f_{\vec p}^{(0)}
\eea
where $I_{\pm}$ denote the even-$m$ and odd-$m$ parts of $I_{\rm ee}$. 
Since $I_+=-\gamma$, the first equation yields a relation 
$\delta f_{\vec p}^{+}(t,\vec x)
=-\frac1{\gamma}\vec v\nabla_{\vec x}\delta f_{\vec p}^{-}(t,\vec x)$, valid at low frequencies $\omega\ll\gamma$, i.e. at the 
lengthscales $r\gg l_{\rm ee}$.
Plugging it 
in the second equation and interpreting $I_-$ as the angle diffusion operator,
\be\label{eq: angle_diffusion}
I_-=\sum_{m\ {\rm odd}}-\gamma_m \left.|m\ra\la m|\right.\approx -\gamma'(i\p_\theta)^p
\ee
yields a closed-form relation for $\delta f_{\vec p}^{-}$ that will serve as a master equation for the new transport regime
\be\label{eq:master_eqn}
\lb \p_t -  
D(\hat{\vec v}\nabla_{\vec x})^2+\gamma'(i\p_\theta)^p\rb \delta f_{\vec p}^{-}(t,\vec x)
=-e\vec E(\vec x)\nabla_{\vec p}f_{\vec p}^{(0)}
,
\ee
where we defined $D=v^2/\gamma$. 
Eq.\eqref{eq:master_eqn} describes ``tomographic dynamics'': fast one-dimensional spatial diffusion along 
unchanging direction 
of velocity $\vec v$ accompanied by a slow angle diffusion that gradually randomizes the orientation of $\vec v$. 

In the above derivation we ignored the $m=0$ zero mode of $I_+$ since in the shear flows created by  transverse fields $\vec E_{\vec k}\perp\vec k$ particle density remains unperturbed. An extension of Eq.\eqref{eq:master_eqn} accounting for this mode will be discussed elsewhere. 
Zero modes of $I_-$ with $m=\pm 1$ can be accounted for by replacing in Eqs.\eqref{eq: angle_diffusion},\eqref{eq:master_eqn} $\p_\theta^2\to \p_\theta^2-1$. 
However, this change only matters in the long-wavelength hydrodynamic regime, at $r\gtrsim\xi$, but 
would not affect the behavior in the 
tomographic regime, Eq.\eqref{eq:lengthscales}. We therefore suppress such terms for the time being. 

A perturbed momentum distribution can be obtained by inverting transport operator in Eq.\eqref{eq:master_eqn}.
Passing to Fourier representation $\delta f_{\vec p}(t,\vec x)=\delta f_{\vec p}e^{-i\omega t+i\vec k\vec x}$ we write a formal operator solution of Eq.\eqref{eq:master_eqn} as 
\be\label{eq:general_soln}
\delta f_{\vec p}=
-\frac1{\hat L-i\omega}
e\vec E\nabla_{\vec p}f_{\vec p}^{(0)}
,\quad
\hat L= 
D(\hat{\vec v}\vec k)^2
+\gamma'(i\p_\theta)^p 
.
\ee
Writing $\vec E\nabla_{\vec p}f_{\vec p}^{(0)}=\vec E\vec v
\frac{\p f_{\vec p}^{(0)}}{\p\epsilon}$ 
and noting that $-\frac{\p f_{\vec p}^{(0)}}{\p\epsilon}=\beta f_{\vec p}^{(0)}(1-f_{\vec p}^{(0)})\approx \delta(\epsilon-\mu)$, we see that the resulting 
perturbation indeed peaks at the Fermi level. 
Shear flows arise when $\vec E_{\vec k}=\int d^2 x e^{-i\vec k\vec x}\vec E(\vec x)$ is transverse to $\vec k$; without loss of generality here we take $\vec E_{\vec k}\parallel \hat {\vec y}$, $\vec k\parallel\hat{\vec x}$.

The transport operator $\hat L$ acts on the Fermi surface parameterized by the angle $\theta$; it is a sum of two noncommuting contributions, $(\hat{\vec v}\vec k)^2=k^2\cos^2\theta$ and $(i\p_\theta)^p$. One is diagonal in the $\theta$-representation, the other is diagonal in the $\delta f_m$ representation. 
Diagonalizing $\hat L$, therefore,
represents a nontrivial task.  
Assuming that the eigenfunctions and eigenvalues of $\hat L$, defined by $\hat L \psi_n(\theta)=\lambda_n\psi_n(\theta)$, are known, we can write the inverse as
\be\label{eq:L_inverse}
\la\theta\Big|\frac1{\hat L-i\omega} 
\Big|\theta'\ra
=\sum_n\frac{\bar\psi_n(\theta)\psi_n(\theta')}{\lambda_n-i\omega}
\ee 
Using 
Eq.\eqref{eq:L_inverse} we 
proceed to evaluate current $j_{y,\vec k}=e v\nu_0 \oint \frac{d\theta}{2\pi}\sin\theta \delta f(\theta)$, where $\nu_0$ is the density of states at $\epsilon_F$. 
Plugging the angle dependence $\vec E\vec v=Ev\sin\theta$ gives 
\be\label{eq:currentPP}
\vec j_{\vec k}=e^2v^2\nu_0 \vec E_{\vec k} 
\oint \frac{d\theta}{2\pi}\oint \frac{d\theta'}{2\pi}  \sin\theta\la\theta\Big|\frac1{\hat L-i\omega} 
\Big|\theta'\ra \sin\theta' 
.
\ee
We can rewrite this relation as $\vec j_{\vec k}=\sigma(k,\omega)\vec E_{\vec k}$ by introducing a scale dependent conductivity 
\be\label{eq:sigma(w,k)}
\sigma(k,\omega)=e^2v^2\nu_0\sum_n\frac{|\la\sin\theta|\psi_n(\theta)\ra|^2}{\lambda_n-i\omega}
\ee
The matrix elements $\la\sin\theta|\psi_n(\theta)\ra$ quickly decrease with $n$, allowing to estimate the sum in Eq.\eqref{eq:sigma(w,k)} by retaining only the $n=0$ term. The lowest eigenvalue can be found by the variational method as
\be\label{eq:variational}
\lambda_0={\rm min}\la \psi|\hat L
|\psi\ra 
\sim {\rm min} \lp Dk^2\delta\theta^2+\frac{\gamma'}{\delta\theta^p}\rp 
\ee
Here the trial state is normalized, $\la \psi|\psi\ra=1$, and is localized within the region of width $\delta\theta$ near the minima of $\cos^2\theta$, i.e. around $\theta =\pm\pi/2$. The estimate in Eq.\eqref{eq:variational} gives the width $\delta\theta\sim (\gamma'/Dk^2)^{\frac1{2+p}}$ and the value
\be
\lambda_0\sim Dk^2\lp\frac{\gamma'}{Dk^2}\rp^{\frac2{2+p}}
\ee
Plugging these values in Eq.\eqref{eq:sigma(w,k)} and setting $\omega=0$, gives a scale-dependent DC conductivity
\be\label{eq:sigma(k)_scaling}
\sigma(k)\sim \frac{e^2v^2\nu_0}{Dk^2}\lp\frac{Dk^2}{\gamma'}\rp^{\frac1{2+p}}\sim k^{-2+\frac{2}{2+p}}
\ee
The variational estimate that leads to this answer is valid provided $\delta\theta\ll 1$, which translates into the condition $k>(\gamma'/D)^{1/2}=1/\xi$ identical to the upper limit in Eq.\eqref{eq:lengthscales} which marks the tomographic-hydrodynamic crossover. 

Viscosity scale dependence can now be inferred by comparing 
Eq.\eqref{eq:sigma(k)_scaling} to the 
conductivity $\sigma(k)=\frac{n^2e^2}{\eta k^2}$ obtained from the Stokes equation $-\eta\nabla^2\vec v=ne\vec E$, giving
\be\label{eq:eta(k)}
\eta(k)\sim k^{-\frac{2}{2+p}}
,
\ee
Eq.\eqref{eq:eta(k)} predicts 
viscosity growing vs. lengthscale, in agreement with the qualitative picture discussed above. The scaling exponents are $-1/3$, $-1/2$ and $-1$ for the three cases $p=4,\,2,\,0$ discussed beneath Eq.\eqref{eq:I_m}.

These results are valid for wavenumbers in the range 
$l_{\rm ee}^{-1}> k> \xi^{-1}$, 
see Eq.\eqref{eq:lengthscales}. Larger values $k> l_{\rm ee}^{-1}$ correspond to ballistic free-particle transport; smaller values $k<\xi^{-1}$ correspond to hydrodynamic transport. 
At $k\xi \sim 1$ our $k$-dependent viscosity values $\nu(k)$ match the standard hydrodynamic value $\eta_{\rm hydro}=nmv^2/4\gamma$. 
At shorter lengthscales, $k\xi >1$, the viscosity is reduced compared to $\eta_{\rm hydro}$ by a factor $(k\xi)^{\frac2{2+p}}$. The reduction in $\eta$ is maximal at 
$k\sim l_{\rm ee}^{-1}$, where $\eta(k)/\eta_{\rm hydro}\sim(\gamma'/\gamma)^{\frac1{2+p}}$. This scale dependence implies that, somewhat 
unexpectedly, the system behavior is more fluid-like at smaller distances and more gaseous at larger distances. 

\begin{figure}
\includegraphics[scale=0.19]{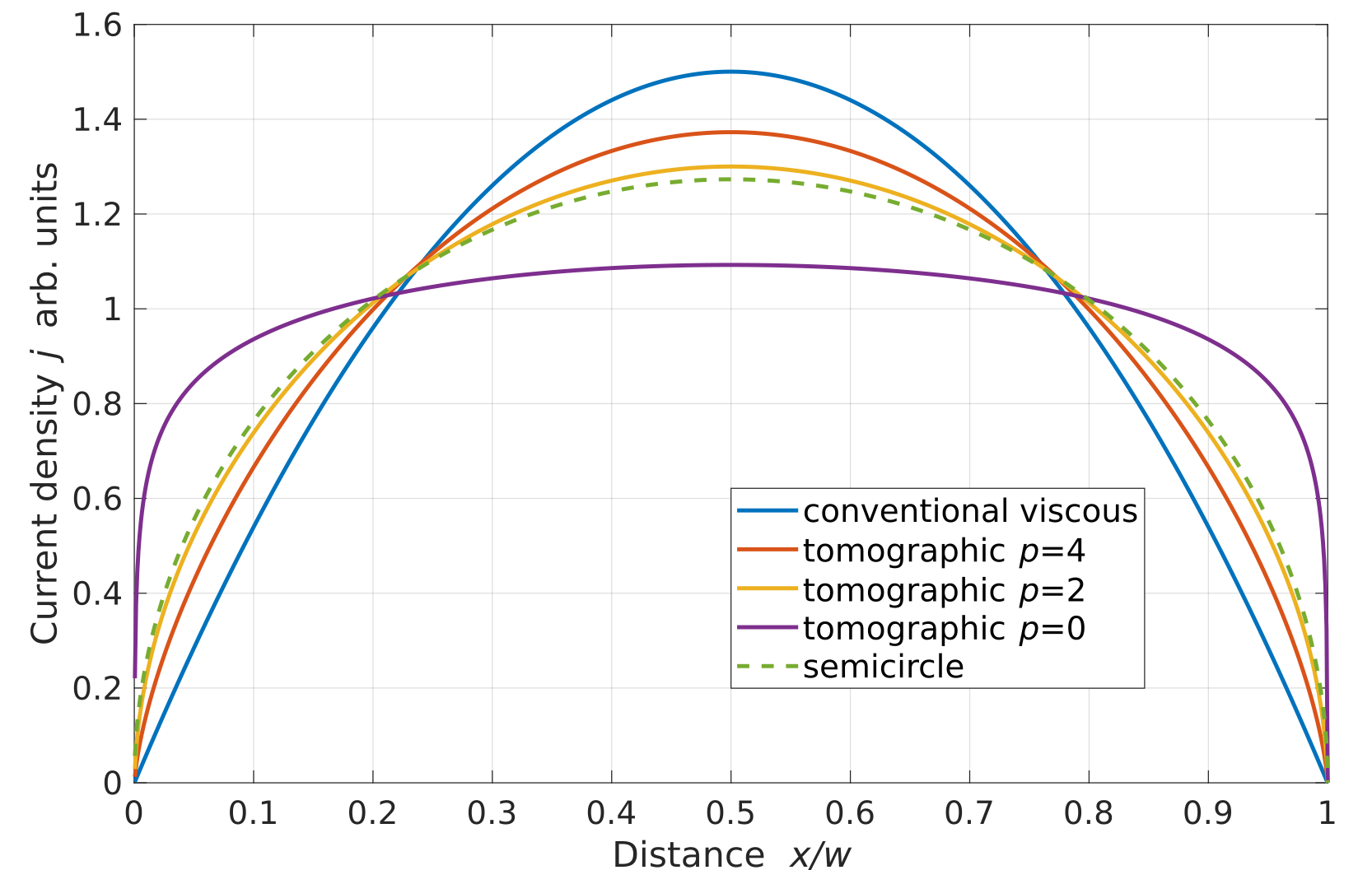} 
\caption{ 
Current density in a long strip 
of width $w$ induced by a uniform DC electric field, Eq.\eqref{eq:j(x)}. The flow profile is different for the viscous and tomographic regimes, showing signatures that depend on the angular relaxation dynamics type, parameterized by $p=4,\,2,\,0$. A semicircle is shown as a guide to the eye. 
}
\label{fig2}
\vspace{-5mm}
\end{figure}

Next, we demonstrate that scale dependence of $\sigma$ and $\eta$ manifests itself in a characteristic current distribution across sample crosssection, which is distinct from the familiar parabolic distribution for conventional viscous flows. 
We analyze flow in a strip $0<x<w$, $-\infty<y<\infty$ with momentum relaxation at the boundaries $x=0,w$. To simplify the geometry, 
we consider an auxiliary problem in an infinite $(x,y)$ plane 
equipped with an array of lines, spaced by $w$, where current relaxation may occur. Current induced by an $E$ field, which is parallel to the lines, is given by
\be\label{eq:j_E}
j(x) =\int dx'\sigma(x-x')\lb E-\alpha \textstyle{\sum_i} j(x_i)\delta(x-x_i)\rb
,
\ee
with $x_i=wi$. Here $\alpha$ is a parameter that is a property of the lines, representing strip boundary, and $\sigma(x-x')=\int \frac{dk}{2\pi} e^{ik(x-x')}\sigma(k)$. The limit of interest to us is $\alpha\to\infty$.

Current distribution for this problem can be obtained by the Fourier method, by writing 
\be
j(x)=\sum_n j_n e^{k_nx}
,\quad
k_n=\frac{2\pi}{w}n
,\quad
n=0,\pm1,\pm2...
\ee 
Plugging this expression in Eq.\eqref{eq:j_E} and Fourier transforming, we have a system of coupled equations for $j_n$:
\be
\rho_n j_n=E\delta_{n,0}-\tilde\alpha\sum_{n'}j_{n'}
,\quad
\rho_n=\frac1{\sigma(k_n)}
,
\ee
where we defined $\tilde\alpha=\frac{\alpha}{w}$. 
These equations can be solved by separating the $n=0$ and $n\ne 0$ harmonics,
\be\label{eq:m_zero_nonzero}
(\rho_0+\tilde\alpha)j_0=E-\tilde\alpha \textstyle{\sum'} j_{n'}
, \quad
 j_{n}=\sigma(k_n)\lp -\tilde\alpha j_0-\tilde\alpha \textstyle{\sum'} j_{n'}\rp
.
\ee
where we introduced a shorthand notation $\textstyle{\sum'}=\sum_{n'\ne 0}$. 
Taking a sum over all $n\ne0$ harmonics yields a relation
\be
(1+\tilde\alpha G)\textstyle{\sum'} j_{n}=-\tilde\alpha G j_0
,\quad
G=\textstyle{\sum'}\sigma(k_n)
.
\ee
Expressing $\textstyle{\sum'} j_{m'}$ and combining with the first equation in Eq.\eqref{eq:m_zero_nonzero}, we obtain
\be
\lp \rho_0+\frac{\tilde\alpha}{1+\tilde\alpha G}\rp j_0=E
\ee
For the case when there are no ohmic losses, $\rho_0=0$, and in the limit $\alpha\to \infty$, this relation simplifies to
\be\label{eq:j0}
j_0=
E\, \textstyle{\sum'}\sigma(k_n)
.
\ee
The distribution of current within the strip then is
\be\label{eq:j(x)}
j(x)=j_0 \lp 1-\frac{\textstyle{\sum'}\sigma(k_n) e^{ik_nx}}{
\textstyle{\sum'}\sigma(k_n)}\rp
.
\ee
For conventional scale-independent viscosity, plugging $\sigma(k)=\frac1{\nu k^2}$, this expression, after a little algebra, gives the familiar parabolic profile $j(x)\sim x(w-x)$. For scale-dependent viscosity $\nu(k)\sim k^{-\frac{2}{2+p}}$ it yields a distribution closely resembling the fractional-power profile 
\be
j(0<x<w)\sim x^{\frac{2}{2+p}}(w-x)^{\frac{2}{2+p}}
.
\ee
The resulting current profiles are illustrated in Fig.\ref{fig2} for several cases of interest. We see that the $k$ dependence of $\sigma$ and $\eta$ has a strong impact on the current profile, providing a directly measurable signature of the tomographic regime. 

This analysis points to several other interesting aspects of tomographic dynamics. First, the system conductance dependence vs.  strip width can be obtained by noting that the sum in Eq.\eqref{eq:j0} converges rapidly, and is well approximated by the
first term, $m = 1$. This predicts scaling for the conductance of the form
\be
G(w)\sim w^{3-\frac2{2+p}}
,
\ee 
a dependence 
that {\it lies in between} the seminal  Gurzhi scaling $w^3$ for the conventional viscous regime\cite{gurzhi63} and $w^2$ scaling for the ballistic transport regime\cite{beenakker_91}. 

Second, velocities of current-carrying electrons are tightly collimated along the strip axis, 
spanning angles in the range  estimated above, $\delta\theta_{k_1}=(\xi k_1)^{-\frac2{2+p}}\ll 1$.
This is in stark contrast to 
conventional viscous flows, 
where velocities are nearly isotropic. 
Strong velocity collimation tunable by
the ee collision rate  
is a surprising behavior, which, along with the peculiar fractional-power conductance scaling, provides a clear signature of 
the tomographic regime.

Part of this work was performed at the Aspen Center for Physics, which is supported by National Science Foundation grant PHY-1607611. We acknowledge support by the MIT Center for Excitonics, the Energy Frontier Research Center funded by the US Department of Energy, Office of Science under Award de-sc0001088, and Army Research Office Grant W911NF-18-1-0116 (L.L.).

%



\end{document}